\documentclass{elsart}

\usepackage{amssymb,amsmath}

\newcommand\rf[1]{(\ref{eq:#1})}
\newcommand\lab[1]{\label{eq:#1}}
\newcommand\nonu{\nonumber}
\newcommand\br{\begin{eqnarray}}
\newcommand\er{\end{eqnarray}}
\newcommand\be{\begin{equation}}
\newcommand\ee{\end{equation}}

\newcommand\lb{\lbrack}
\newcommand\rb{\rbrack}
\newcommand\llangle{\left\langle}
\newcommand\rrangle{\right\rangle}

\newcommand\llb{\left\lbrack}
\newcommand\rrb{\right\rbrack}

\newcommand\lcurl{\left\{}
\newcommand\rcurl{\right\}}
\renewcommand\({\left(}
\renewcommand\){\right)}
\newcommand\bv{\bigm\vert}               
\newcommand\bgv{\bigg\vert}              

\newcommand\bc{\begin{center}}
\newcommand\ec{\end{center}}

\newcommand\partder[2]{\frac{{\partial {#1}}}{{\partial {#2}}}}

\renewcommand\d{\delta}
\newcommand\D{\Delta}

\newcommand\vareps{\varepsilon}
\newcommand\g{\gamma}
\newcommand\G{\Gamma}

\newcommand\h{\frac{1}{2}}
\renewcommand\k{\kappa}
\renewcommand\l{\lambda}

\newcommand\m{\mu}
\newcommand\n{\nu}

\newcommand\om{\omega}

\newcommand\vp{\varphi}
\renewcommand\P{\Phi}
\newcommand\pa{\partial}

\newcommand\pr{\prime}

\renewcommand\r{\rho}
\newcommand\s{\sigma}
\renewcommand\S{\Sigma}
\renewcommand\t{\tau}
\renewcommand\th{\theta}

\newcommand\wti{\widetilde}
\newcommand\twomat[4]{\left(\begin{array}{cc}  
{#1} & {#2} \\ {#3} & {#4} \end{array} \right)}

\newcommand\cA{{\mathcal A}}

\newcommand\cF{{\mathcal F}}

\newcommand{\ct}[1]{\cite{#1}}
\newcommand{\bib}[1]{\bibitem{#1}}

\newcommand\PRL[3]{\textsl{Phys. Rev. Lett.} \textbf{#1} (#2) #3}

\newcommand\PRD[3]{\textsl{Phys. Rev.} \textbf{D#1} (#2) #3}

\newcommand\CQG[3]{\textsl{Class. Quantum Grav.} \textbf{#1} (#2) #3}

\newcommand\AoP[3]{\textsl{Ann. of Phys.} \textbf{#1} (#2) #3}

\newcommand\IJMPA[3]{\textsl{Int. J. Mod. Phys.} \textbf{A#1} (#2) #3}

\newcommand\MPLA[3]{\textsl{Mod. Phys. Lett.} \textbf{A#1} (#2) #3}

\newcommand\Xdot{\stackrel{.}{X}}

\newcommand\rdot{\stackrel{.}{r}}

\newcommand\vpdot{\stackrel{.}{\varphi}}

\begin{document}

\begin{frontmatter}



\title{Lightlike Brane as a Gravitational Source \\ of Misner-Wheeler-Type Wormholes}


\author[BGU]{Eduardo Guendelman\corauthref{cor1}},
\ead{guendel@bgu.ac.il}
\corauth[cor1]{Corresponding author -- tel. +972-8-647-2508, fax +972-8-647-2904.}
\author[BGU]{Alexander Kaganovich\corauthref{cor1}}
\ead{alexk@bgu.ac.il}
\address[BGU]{Department of Physics, Ben-Gurion University of the Negev, 
P.O.Box 653, IL-84105 ~Beer-Sheva, Israel}

\author[INRNE]{Emil Nissimov\thanksref{RTN-F1412}},
\ead{nissimov@inrne.bas.bg}
\author[INRNE]{Svetlana Pacheva\thanksref{RTN-F1412}}
\thanks[RTN-F1412]{Supported by European RTN network
{\em ``Forces-Universe''} (contract No.\textsl{MRTN-CT-2004-005104}).}
\ead{svetlana@inrne.bas.bg}
\address[INRNE]{Institute for Nuclear Research and Nuclear Energy, Bulgarian Academy
of Sciences, Boul. Tsarigradsko Chausee 72, BG-1784 ~Sofia, Bulgaria}

\begin{abstract}
Consistent Lagrangian description of lightlike $p$-branes (\textsl{LL-branes})
is presented in two equivalent forms -- a Polyakov-type formulation and a dual 
to it Nambu-Goto-type formulation. An important characteristic feature of the
\textsl{LL-branes} is that the brane tension appears as a non-trivial
additional dynamical degree of freedom. Next, properties of $p=2$ \textsl{LL-brane}
dynamics (as a test brane) in $D=4$ Kerr or Kerr-Newman gravitational backgrounds 
are discussed in some detail. It is shown that the \textsl{LL-brane} automatically 
positions itself on the horizon and rotates along with the same angular velocity.
Finally, a Misner-Wheeler-type of Reissner-Nordstr{\"o}m wormhole is constructed in a
self-consistent electrically sourceless Einstein-Maxwell system in the $D=4$ bulk 
interacting with a \textsl{LL-brane}. The pertinent wormhole throat is located 
precisely at the \textsl{LL-brane} sitting on the outer Reissner-Nordstr{\"o}m
horizon with the Reissner-Nordstr{\"o}m mass and charge being functions
of the dynamical \textsl{LL-brane} tension.
\end{abstract}

\end{frontmatter}

\section{Introduction}
\label{intro}
Lightlike branes (\textsl{LL-branes}, for short) attract special interest in 
general relativity. This is due primarily because of their role in the
effective description of many cosmological and astrophysical effects:
(a) impulsive lightlike signals arising in cataclysmic astrophysical events
\ct{barrabes-hogan}; 
(b) the ``membrane paradigm'' theory of black hole physics 
\ct{membrane-paradigm}; 
(c) thin-wall approach to domain walls coupled to gravity
\ct{Israel-66,Barrabes-Israel-Hooft}.
More recently \textsl{LL-branes} acquired significance also in the context of
modern non-perturbative string theory \ct{nonperturb-string}.

Our formalism makes an essential use of an alternative non-Riemannian measure of 
integration (volume-form). The latter leads to different type of gravitational 
theories \ct{TMT} which address various basic problems of 
cosmological interest. In the context of the theory of extended
objects employing an alternative integration measure independent of the
intrinsic Riemannian metric on the world-volume within the Polyakov-type
approach leads to a dynamical string/brane tension \ct{mod-measure}. Furthermore it
allows the construction of consistent Lagrangian actions describing
intrinsically lightlike $p$-branes (\textsl{LL-branes}) \ct{LL-branes-1}.
Also an equivalent Nambu-Goto-type formulation of \textsl{LL-brane} dynamics
has been shown to exist (third ref.\ct{LL-branes-2}, cf. Eq.\rf{LL-action-NG} below) 
which is dual to the Polyakov-type formulation (cf. Eq.\rf{LL-brane} below).

In a series of papers \ct{LL-branes-1,LL-branes-2} we have studied the properties of
\textsl{LL-branes} both as test branes moving in physically interesting
gravitational backgrounds, as well as material and charge sources for
gravity and electromagnetism in self-consistent bulk gravity-matter systems
interacting with \textsl{LL-branes}.

In gravitational backgrounds of spherically symmetric type and codimension-one
a general feature of \textsl{LL-branes} is that their dynamics is consistent 
only provided the background possesses an event horizon which is automatically 
occupied by the \textsl{LL-brane}. Also, the dynamical brane tension
exhibits an exponential ``inflation/deflation'' property analogous to the
``mass inflation effect'' around black hole horizons discovered in
\ct{poisson-israel}. Furthermore, unlike conventional braneworlds, where the 
underlying branes are of Nambu-Goto type and in their ground state they position 
themselves at some fixed point in the extra dimensions of the bulk space-time, 
codimension-two (or more) {\em lightlike braneworlds} perform in the ground state 
non-trivial motions in the extra dimensions -- planar circular, spiral winding 
\textsl{etc} depending on the topology of the extra dimensions.
For details we refer to \ct{LL-branes-1,LL-branes-2}.

In the present note we are going to study dynamics of \textsl{LL-branes}
both as test branes and material sources in the case of Kerr-Newman black
hole space-time, in particular -- Reissner-Nordstr{\"o}m space-time as a
limiting case of the former (cf. the textbooks \ct{textbooks-kerr}). 
We find that the \textsl{LL-brane} automatically positions itself on the
Kerr-Newman horizon and in addition it rotates along with the same angular
velocity as the black hole. Further, we find a self-consistent 
Einstein-Maxwell-\textsl{LL-brane} solution where one ``surgically'' eliminates
the space-time region inside the (outer) Reissner-Nordstr{\"o}m horizon via 
sewing together through the \textsl{LL-brane}'s energy-momentum tensor 
(derived from the pertinent \textsl{LL-brane} world-volume action) two copies of 
the exterior Reissner-Nordstr{\"o}m region along their common (outer) horizon 
such that the normal radial-like coordinate decreases when we approach the 
\textsl{LL-brane} sitting on the horizon from one side and increases as we
continue to the other side. In other words, one achieves a Reissner-Nordstr{\"o}m
wormhole solution of Misner-Wheeler type \ct{Misner-Wheeler} which does not require
electrically charged sources.

Let us particularly stress that our construction below is based on a first
principle's approach, \textsl{i.e.}, the ``surgical'' matching at the
Reissner-Nordstr{\"o}m wormhole ``throat'' comes from a well-defined world-volume
Lagrangian description of \textsl{LL-brane} dynamics.

\section{World-Volume Lagrangian Description of Lightlike Branes}
\label{action}
In refs.\ct{LL-branes-1,LL-branes-2} we have proposed a systematic Lagrangian 
formulation of a generalized Polyakov-type for \textsl{LL-branes} in terms of the 
world-volume action:
\be
S_{\mathrm{LL}} = \int d^{p+1}\s \,\P (\vp) \llb - \h \g^{ab} g_{ab} + 
L\!\( F^2\)\rrb \; ,
\lab{LL-brane}
\ee
with the following notations. Here $\g_{ab}$ denotes the intrinsic Riemannian 
metric on the world-volume, $a,b=0,1,\ldots ,p$; $(\s^a)\equiv (\t,\s^i)$ with
$i=1,\ldots ,p$; $g_{ab}$ is the induced metric:
\be
g_{ab} \equiv \pa_a X^{\m} \pa_b X^{\n} G_{\m\n}(X) \; ,
\lab{ind-metric}
\ee
which becomes {\em singular} on-shell (manifestation of the lightlike nature), 
cf. second Eq.\rf{phi-gamma-eqs} below);
\be
\P (\vp) \equiv \frac{1}{(p+1)!} \vareps_{I_1\ldots I_{p+1}}
\vareps^{a_1\ldots a_{p+1}} \pa_{a_1} \vp^{I_1}\ldots \pa_{a_{p+1}} \vp^{I_{p+1}}
\lab{mod-measure-p}
\ee
is an alternative non-Riemannian reparametrization-covariant integration measure 
density replacing the standard $\sqrt{-\g} \equiv \sqrt{-\det \Vert \g_{ab}\Vert}$
and built from auxiliary world-volume scalars $\lcurl \vp^I \rcurl_{I=1}^{p+1}$;
\be
F^2 \equiv F_{a_1 \ldots a_{p}} F_{b_1 \ldots b_{p}} 
\g^{a_1 b_1} \ldots \g^{a_p b_p} \; ,
\nonu
\ee
where:
\be
F_{a_1 \ldots a_{p}} = p \pa_{[a_1} A_{a_2\ldots a_{p}]} \quad ,\quad
F^{\ast a} = \frac{1}{p!} \frac{\vareps^{a a_1\ldots a_p}}{\sqrt{-\g}}
F_{a_1 \ldots a_{p}}
\lab{p-rank}
\ee
are the field-strength and its dual one of an auxiliary world-volume $(p-1)$-rank 
antisymmetric tensor gauge field $A_{a_1\ldots a_{p-1}}$ with Lagrangian $L(F^2)$. 

Equivalently one can rewrite \rf{LL-brane} as:
\be
S_{\mathrm{LL}} = \int d^{p+1}\!\!\s \,\chi \sqrt{-\g} 
\llb -\h \g^{ab} g_{ab} + L\!\( F^2\)\rrb \quad, \;\; 
\chi \equiv \frac{\P (\vp)}{\sqrt{-\g}}
\lab{LL-brane-chi}
\ee
where from we see that the composite field $\chi$ plays the role of a 
{\em dynamical (variable) brane tension}.

\textbf{Remark.} For the special choice $L\!\( F^2\)= \( F^2\)^{1/p}$ the 
above action becomes invariant under Weyl (conformal) symmetry: 
\be
\g_{ab} \longrightarrow \g^{\pr}_{ab} = \rho\,\g_{ab}  \quad ,\quad
\vp^{i} \longrightarrow \vp^{\pr\, i} = \vp^{\pr\, i} (\vp)
\lab{Weyl-conf}
\ee
with Jacobian $\det \Bigl\Vert \frac{\pa\vp^{\pr\, i}}{\pa\vp^j} \Bigr\Vert = \rho$. 

Now let us consider the equations of motion corresponding to \rf{LL-brane} 
w.r.t. $\vp^I$ and $\g^{ab}$:
\be
\h \g^{cd} g_{cd} - L(F^2) = M \quad , \quad
\h g_{ab} - F^2 L^{\pr}(F^2) \llb\g_{ab} 
- \frac{F^{*}_a F^{*}_b}{F^{*\, 2}}\rrb = 0  \; .
\lab{phi-gamma-eqs}
\ee
Here $M$ is an integration constant and $F^{*\, a}$ is the dual field
strength \rf{p-rank}. Both Eqs.\rf{phi-gamma-eqs} imply the constraint:
\be
L\!\( F^2\) - p F^2 L^\pr\!\( F^2\) + M = 0,\; \textsl{i.e.}\;
F^2 = F^2 (M) = \mathrm{const} ~\mathrm{on-shell} \; .
\lab{F2-const}
\ee
The second Eq.\rf{phi-gamma-eqs} exhibits {\em on-shell singularity} 
of the induced metric \rf{ind-metric}:
\be
g_{ab}F^{*\, b} \equiv \pa_a X^\m G_{\m\n} \( F^{*\, b}\pa_b X^\n\) =0 \; ,
\lab{on-shell-singular}
\ee
\textsl{i.e.}, the tangent vector to the world-volume $F^{*\, a}\pa_a X^\m$
is lightlike w.r.t. metric of the embedding space-time.

Further, the equations of motion w.r.t. world-volume gauge field 
$A_{a_1\ldots a_{p-1}}$ (with $\chi$ as defined in \rf{LL-brane-chi} and
accounting for the constraint \rf{F2-const}) read:
\be
\pa_{[a}\( F^{\ast}_{b]}\, \chi\) = 0  \; .
\lab{A-eqs}
\ee
They allow us to introduce the dual ``gauge'' potential $u$:
\be
F^{\ast}_{a} = \mathrm{const}\, \frac{1}{\chi} \pa_a u \; ,
\lab{u-def}
\ee
enabling us to rewrite second Eq.\rf{phi-gamma-eqs} (the lightlike constraint)
in terms of the dual potential $u$ in the form:
\be
\g_{ab} = \frac{1}{2a_0}\, g_{ab} - \frac{2}{\chi^2}\,\pa_a u \pa_b u \quad ,
\quad a_0 \equiv F^2 L^{\pr}\( F^2\)\bv_{F^2=F^2(M)} = \mathrm{const}
\lab{gamma-eqs-u}
\ee
($L^\pr(F^2)$ denotes derivative of $L(F^2)$ w.r.t. the argument $F^2$).
From \rf{u-def} and \rf{F2-const} we obtain the relation: 
\be
\chi^2 = -2 \g^{ab} \pa_a u \pa_b u \; ,
\lab{chi2-eq}
\ee
and the Bianchi identity $\nabla_a F^{\ast\, a}=0$ becomes:
\be
\pa_a \Bigl( \frac{1}{\chi}\sqrt{-\g} \g^{ab}\pa_b u\Bigr) = 0  \; .
\lab{Bianchi-id}
\ee

Finally, the $X^\m$ equations of motion produced by the \rf{LL-brane} read:
\be
\pa_a \(\chi \sqrt{-\g} \g^{ab} \pa_b X^\m\) + 
\chi \sqrt{-\g} \g^{ab} \pa_a X^\n \pa_b X^\l \G^\m_{\n\l}(X) = 0  \;
\lab{X-eqs}
\ee
where $\G^\m_{\n\l}=\h G^{\m\k}\(\pa_\n G_{\k\l}+\pa_\l G_{\k\n}-\pa_\k G_{\n\l}\)$
is the Christoffel connection for the external metric.

Now it is straightforward to prove that the system of equations 
\rf{chi2-eq}--\rf{X-eqs} for $\( X^\m,u,\chi\)$, which are equivalent to the 
equations of motion \rf{phi-gamma-eqs}--\rf{A-eqs},\rf{X-eqs} resulting from the 
original Polyakov-type \textsl{LL-brane} action \rf{LL-brane}, can be equivalently 
derived from the following {\em dual} Nambu-Goto-type world-volume action: 
\be
S_{\rm NG} = - \int d^{p+1}\s \, T 
\sqrt{- \det\Vert g_{ab} - \frac{1}{T^2}\pa_a u \pa_b u\Vert}  \; .
\lab{LL-action-NG}
\ee
Here $g_{ab}$ is the induced metric \rf{ind-metric};
$T$ is {\em dynamical} tension simply related to the dynamical tension 
$\chi$ from the Polyakov-type formulation \rf{LL-brane-chi} as
$T^2= \frac{\chi^2}{4a_0}$ with $a_0$ -- same constant as in \rf{gamma-eqs-u}.

In what follows we will consider the initial Polyakov-type form \rf{LL-brane} of the
\textsl{LL-brane} world-volume action. World-volume reparametrization invariance 
allows to introduce the standard synchronous gauge-fixing conditions:
\be
\g^{0i} = 0 \;\; (i=1,\ldots,p) \; ,\; \g^{00} = -1
\lab{gauge-fix}
\ee
Also, we will use a natural ansatz for the ``electric'' part of the 
auxiliary world-volume gauge field-strength:
\be
F^{\ast i}= 0 \;\; (i=1,{\ldots},p) \quad ,\quad \mathrm{i.e.} \;\;
F_{0 i_1 \ldots i_{p-1}} = 0 \; ,
\lab{F-ansatz}
\ee
meaning that we choose the lightlike direction in Eq.\rf{on-shell-singular} 
to coincide with the brane
proper-time direction on the world-volume ($F^{*\, a}\pa_a \simeq \pa_\t$).
The Bianchi identity ($\nabla_a F^{\ast\, a}=0$) together with 
\rf{gauge-fix}--\rf{F-ansatz} and the definition for the dual field-strength
in \rf{p-rank} imply:
\be
\pa_0 \g^{(p)} = 0 \quad \mathrm{where}\;\; \g^{(p)} \equiv \det\Vert\g_{ij}\Vert \; .
\lab{gamma-p-0}
\ee
Then \textsl{LL-brane} equations of motion acquire the form 
(recall definition of $g_{ab}$ \rf{ind-metric}):
\be
g_{00}\equiv \Xdot^\m\!\! G_{\m\n}\!\! \Xdot^\n = 0 \quad ,\quad g_{0i} = 0 \quad ,\quad
g_{ij} - 2a_0\, \g_{ij} = 0
\lab{gamma-eqs-0}
\ee
(the latter are analogs of Virasoro constraints), where the $M$-dependent constant
$a_0$ (the same as in \rf{gamma-eqs-u}) must be strictly positive;
\be
\pa_i \chi = 0 \qquad (\mathrm{remnant ~of ~Eq.\rf{A-eqs}})\; ;
\lab{A-eqs-0}
\ee

\vspace{-0.6cm}
\br
-\sqrt{\g^{(p)}} \pa_0 \(\chi \pa_0 X^\m\) +
\pa_i\(\chi\sqrt{\g^{(p)}} \g^{ij} \pa_j X^\m\)
\nonu \\
+ \chi\sqrt{\g^{(p)}} \(-\pa_0 X^\n \pa_0 X^\l + \g^{kl} \pa_k X^\n \pa_l X^\l\)
\G^\m_{\n\l} = 0 \; .
\lab{X-eqs-0}
\er

\section{Lightlike Branes in Kerr-Newman Black Hole Background}
\label{kerr}
Let us consider $D\! =\! 4$-dimensional Kerr-Newman background metric in the 
standard Boyer-Lindquist coordinates (see \textsl{e.g.} \ct{textbooks-kerr}):
\be
ds^2 = -A (dt)^2 - 2 E dt\,d\vp + \frac{\S}{\D} (dr)^2 + \S (d\th)^2 +
D \sin^2 \th (d\vp)^2  \; ,
\lab{kerr-metric}
\ee
\be
A\equiv \frac{\D - a^2 \sin^2 \th}{\S} \;\; ,\;\;
E\equiv \frac{a \sin^2 \th\,\(r^2 + a^2 - \D\)}{\S} \;\; ,\;\;
D\equiv \frac{\( r^2 + a^2\)^2 - \D a^2 \sin^2 \th}{\S} \; ,
\lab{kerr-coeff}
\ee
where $\S \equiv r^2 + a^2\cos^2 \th\; ,\; \D \equiv r^2 + a^2 + e^2 - 2 Mr$.
Let us recall that the Kerr-Newman metric \rf{kerr-metric}--\rf{kerr-coeff}
reduces to the Reissner-Nordstr{\"o}m metric in the limiting case $a=0$.

For the \textsl{LL-brane} embedding we will use the following ansatz:
\be
X^0 \equiv t = \t \;\; ,\;\; r=r(\t) \;\; ,\;\; \th = \s^1 \;\; ,\;\;
\vp = \s^2 + {\wti \vp}(\t) \; .
\lab{kerr-ansatz}
\ee
In this case the \textsl{LL-brane} equations of motion
\rf{gamma-p-0}--\rf{gamma-eqs-0} acquire the form:
\br
-A + \frac{\S}{\D} \rdot^2 + D \sin^2 \th\,\vpdot^2 - 2 E \vpdot = 0
\nonu \\
-E + D \sin^2 \th\,\vpdot = 0 \quad ,\quad  
\frac{d}{d\t}\( D\S \sin^2 \th\) = 0 \; .
\lab{gamma-eqs-kerr}
\er
Inserting the ansatz \rf{kerr-ansatz} into \rf{gamma-eqs-kerr} the last 
Eq.\rf{gamma-eqs-kerr} implies:
\be
r(\t) = r_0 = \mathrm{const} \; , 
\lab{kerr-horizon}
\ee
whereas the second Eq.\rf{gamma-eqs-kerr} yields:
\be
\D (r_0) = 0 \quad ,\quad \om \equiv \vpdot = \frac{a}{r_0^2 + a^2}
\lab{kerr-dragging}
\ee
Eqs.\rf{kerr-horizon}--\rf{kerr-dragging} indicate that:\\
$\phantom{aa}$(i) the \textsl{LL-brane} 
automatically locates itself on the Kerr-Newman horizon $r=r_0$ -- horizon 
``straddling'' according to the terminology of the first 
ref.\ct{Barrabes-Israel-Hooft};\\
$\phantom{aa}$(ii) the \textsl{LL-brane}
rotates along with the same angular velocity $\om$ as the Kerr-Newman horizon.

The first Eq.\rf{gamma-eqs-kerr} implies that $\rdot$ vanishes
on-shell as:
\be
\rdot\, \simeq\, \pm \frac{\D (r)}{r_0^2 + a^2}\bv_{r \to r_0} \; .
\lab{r-dot}
\ee
We will also need the explicit form of the last Eq.\rf{gamma-eqs-0} (using notations 
\rf{kerr-coeff}):
\be
\g_{ij} = \frac{1}{2a_0} \twomat{\S}{0}{0}{D \sin^2 \th}\bgv_{r=r_0 ,\th=\s^1} \; .
\lab{g-ij}
\ee
Among the $X^\m$-equations of motion \rf{X-eqs-0} only the $X^0$-equation
yields additional information. Because of the embedding $X^0 = \t$ it acquires 
the form of a time-evolution equation for the dynamical brane tension $\chi$:
\be
\pa_\t \chi + \chi\,\llb \pa_\t X^\n \pa_\t X^\l 
- \g^{ij} \pa_i X^\n \pa_j X^\l \rrb \G^{0}_{\n\l} = 0 \; ,
\lab{X0-eq}
\ee
which, after taking into account \rf{kerr-ansatz},
\rf{kerr-horizon}--\rf{kerr-dragging}
and the explicit expressions for the Kerr-Newman Christoffel connection 
coefficients (first ref.\ct{textbooks-kerr}), reduces to:
\be
\pa_\t \chi + \chi\, 2\rdot\Bigl\lb \G^0_{0r} + 
\frac{a}{r_0^2 + a^2}\G^0_{r\vp}\Bigr\rb_{r=r_0} = 0 \; .
\lab{X0-eq-1}
\ee
Singularity on the horizon of the Christoffel coefficients $\(\simeq \D^{-1}\)$
appearing in \rf{X0-eq-1} is cancelled by $\D$ in $\rdot$ \rf{r-dot} so that 
finally we obtain:
\be
\pa_\t \chi \pm \chi \frac{2(r_0 - M)}{r_0^2 + a^2} = 0 \;\; ,\; \mathrm{i.e.}\;\;
\chi = \chi_0 \exp\Bigl\{\mp 2 \frac{(r_0 - M)}{r_0^2 + a^2}\,\t\Bigr\}
\lab{chi-eq-kerr}
\ee
Thus, we find ``mass inflation/deflation'' effect (according to the
terminology of \ct{poisson-israel}) on the Kerr-Newman horizon via the exponential time
dependence of the dynamical \textsl{LL-brane} tension. The latter is an analog
of the previously found ``mass inflation/deflation'' effect with \textsl{LL-branes}
in spherically symmetric gravitational backgrounds \ct{LL-branes-2}.

\section{Bulk Einstein-Maxwell System Interacting With a Lightlike Brane and
Misner-Wheeler Wormhole-like Solution}
\label{bulk-sys}
Now we will consider a self-consistent $D=4$ Einstein-Maxwell system coupled to
a \textsl{LL-brane} but free of electrically charged matter, where the 
\textsl{LL-brane} will serve as a gravitational source through its
energy-momentum tensor:
\be
S = \int\!\! d^4 x\,\sqrt{-G}\,\llb \frac{R(G)}{16\pi}
- \frac{1}{4} \cF_{\m\n}\cF^{\m\n}\rrb 
+ S_{\mathrm{LL}} \; .
\lab{E-M-LL}
\ee
Here $\cF_{\m\n} = \pa_\m \cA_\n - \pa_\n \cA_\m$ and $S_{\mathrm{LL}}$ is
the same \textsl{LL-brane} world-volume action as in \rf{LL-brane-chi}.
The pertinent Einstein-Maxwell equations of motion read:
\be
R_{\m\n} - \h G_{\m\n} R =
8\pi \( T^{(EM)}_{\m\n} + T^{(brane)}_{\m\n}\) \quad, \quad
\pa_\n \(\sqrt{-G}G^{\m\k}G^{\n\l} \cF_{\k\l}\) = 0 \; ,
\lab{Einstein-Maxwell-eqs}
\ee
where $T^{(EM)}_{\m\n} = \cF_{\m\k}\cF_{\n\l} G^{\k\l} - G_{\m\n}\frac{1}{4}
\cF_{\r\k}\cF_{\s\l} G^{\r\s}G^{\k\l}$, 
and the \textsl{LL-brane} energy-momentum tensor is straightforwardly derived
from \rf{LL-brane-chi}:
\be
T^{(brane)}_{\m\n} = - G_{\m\k}G_{\n\l}
\int\!\! d^3 \s\, \frac{\d^{(4)}\bigl(x-X(\s)\bigr)}{\sqrt{-G}}\,
\chi\,\sqrt{-\g} \g^{ab}\pa_a X^\k \pa_b X^\l  \; ,
\lab{T-brane}
\ee
Following the standard procedure \ct{Visser-book} we will now construct a 
{\em wormhole} solution to the Einstein-Maxwell equations 
\rf{Einstein-Maxwell-eqs} of Misner-Wheeler type \ct{Misner-Wheeler} by
making an essential use of the explicit expression for the \textsl{LL-brane} 
energy-momentum tensor \rf{T-brane}. Namely, let us take
two copies of Kerr-Newman exterior space-time region, \textsl{i.e.}, solutions to 
\rf{Einstein-Maxwell-eqs} with $G_{\m\n}$ as in \rf{kerr-metric}--\rf{kerr-coeff}
for $r>r_0$, where $r_0 = M + \sqrt{M^2 - a^2 - e^2}$ is the outer horizon
radius, and let us try to sew the two regions together along the horizon 
$r=r_0$ via the \textsl{LL-brane}. To this end it is customary to introduce a 
new radial-like coordinate $\eta$ normal w.r.t. the \textsl{LL-brane}:
\be
r=r_0 + |\eta| \quad ,\quad \partder{r}{\eta}=\mathrm{sign} (\eta) \quad ,\quad
\eta \in \bigl(-\infty,+\infty\bigr) \; .
\lab{r-eta}
\ee
Accordingly, the metric in the total space of the two copies of exterior 
Kerr-Newman regions reads:
\be
ds^2 = - {\wti A} (dt)^2 - 2 {\wti E} dt\,d\vp + \frac{{\wti \S}}{{\wti \D}} (d\eta)^2 
+ {\wti \S} (d\th)^2 + {\wti D} \sin^2 \th (d\vp)^2  \; ,
\lab{kerr-metric-eta}
\ee
where ${\wti A} \equiv A \bv_{r=r_0 + |\eta|}$ with the same $A$ as in 
\rf{kerr-metric}--\rf{kerr-coeff}, and similarly for
${\wti E},\,{\wti \S},\,{\wti \D},\,{\wti D}$. The two copies transform into
each other under the ``parity'' transformation $\eta \to - \eta$.

Inserting in \rf{T-brane} the expressions for $X^\m (\s)$ from
\rf{kerr-ansatz} and \rf{kerr-horizon}--\rf{kerr-dragging}, taking into
account the explicit form of Kerr-Newman metric coefficients 
\rf{kerr-metric}--\rf{kerr-coeff}, the gauge-fixing conditions for the
intrinsic world-volume metric \rf{gauge-fix} and Eq.\rf{g-ij} we get:
\be
T_{(brane)}^{\m\n} = S^{\m\n}\,\d (\eta)
\lab{T-S-0}
\ee
with surface energy-momentum tensor:
\be
S^{\m\n} \equiv - \frac{\chi}{2a_0}\,\frac{r_0^2 + a^2}{r_0^2 + a^2 \cos^2\th}
\llb - \pa_\t X^\m \pa_\t X^\n + \g^{ij} \pa_i X^\m \pa_j X^\n 
\rrb_{t=\t,\,\th =\s^1,\,\vp = \s^2 + \frac{a}{r_0^2 + a^2}\t} \; ,
\lab{T-S-brane}
\ee
where now the indices $\m,\n$ refer to $(t,\eta,\th,\vp)$ and $a_0$ is the integration
constant parameter appearing in the \textsl{LL-brane} dynamics (cf. Eq.\rf{gamma-eqs-u}). 
Let us also note that 
unlike the case ot test \textsl{LL-brane} moving in a Kerr-Newman
background (Eqs.\rf{X0-eq}--\rf{chi-eq-kerr}), the dynamical tension $\chi$ in 
Eq.\rf{T-S-brane} is {\em constant}. This is due to the fact that in the
present context we have a discontinuity in the Kerr-Newman Christoffel connection 
coefficients across the \textsl{LL-brane} sitting on the horizon ($\eta = 0$). 
The latter problem in treating the geodesic \textsl{LL-brane} equations of motion
\rf{X-eqs}, in particular -- Eq.\rf{X0-eq}, is resolved following the approach 
in ref.\ct{Israel-66} (see also the regularization approach in ref.\ct{BGG},
Appendix A) by taking the mean value of the pertinent non-zero Christoffel 
coefficients across the discontinuity at $\eta = 0$ and accounting for \rf{r-eta}:
\be
\llangle\G^0_{0\eta}\rrangle \equiv \h \(\G^0_{0\eta}\bv_{\eta \to +0} +
\G^0_{0\eta}\bv_{\eta \to -0}\) = 
\h \(\G^0_{0r}\bv_{r \to r_0} - \G^0_{0r}\bv_{r \to r_0}\) = 0
\lab{mean-value}
\ee
and similarly for $\llangle\G^0_{\eta\vp}\rrangle = 0$. Therefore, in the latter
case Eq.\rf{X0-eq} is reduced to $\pa_\t \chi = 0$.

From the Einstein equations \rf{Einstein-Maxwell-eqs}, taking into account
Eqs.\rf{T-S-0}--\rf{T-S-brane}, one obtains in a standard way 
the discontinuity for the Kerr-Newman Christoffel coefficients (analog of Israel 
junction conditions \ct{Israel-66,Barrabes-Israel-Hooft}). Namely, observing
that:
\br
R_{\m\n} \equiv \pa_\eta \G^{\eta}_{\m\n} + \pa_\m \pa_\n \ln \sqrt{-G}
+ \mathrm{non-singular ~terms}
\nonu \\
= 8\pi \( S_{\m\n} - \h G_{\m\n} S^{\l}_{\l}\) \d (\eta) 
+ \mathrm{non-singular ~terms} \; ,
\lab{E-M-eqs}
\er
we find that delta-function singularities are present on both sides for 
$(\m\n)=(\eta \eta)$. For $(\m\n)=(0\,\eta)$ and $(\m\n)=(\eta\,\vp)$ such 
singularities appear only on the r.h.s., and the rest of \rf{E-M-eqs} are 
singularity free.
Consistency of \rf{E-M-eqs} for $(\m\n)=(0\,\eta)$ and $(\m\n)=(\eta\,\vp)$,
\textsl{i.e.}, vanishing of the delta-function singularity on the r.h.s.
requires $a=0$. In other words, consistent wormhole solution with \textsl{LL-brane}
as a ``throat'' may exist only for the limiting case of spherically symmetric 
Reissner-Nordstr{\"o}m geometry. 

It remains to check Eq.\rf{E-M-eqs} for $(\m\n)=(\eta \eta)$. In order to
avoid coordinate singularity on the horizon it is more convenient to
consider the mixed component version of the latter:
\be
R^{\eta}_{\eta} = 8\pi \( S^{\eta}_{\eta} - \h S^{\l}_{\l}\) \d (\eta) 
+ \mathrm{non-singular ~terms} \; .
\lab{E-M-eqs-eta}
\ee
Evaluating the r.h.s. of \rf{E-M-eqs-eta} from \rf{T-S-brane}
with \rf{kerr-ansatz} and \rf{kerr-horizon}--\rf{kerr-dragging} we obtain:
\be
\pa_\eta \( (r_0 + |\eta|)^2 \pa_\eta {\wti A}\) =
- 16\pi\, r_0^2\chi\,\d (\eta)\; ,
\lab{E-M-eqs-eta-0}
\ee
where ${\wti A} = \Bigl( 1 - \frac{2M}{r} + \frac{e^2}{r^2}\Bigr)_{r=r_0 + |\eta|}$
is the Reissner-Nordstr{\"o}m limit of the metric coefficient ${\wti A}$ in
\rf{kerr-metric-eta}. Therefore, the junction condition becomes:
\be
\pa_\eta {\wti A}\bv_{\eta \to +0} - \pa_\eta {\wti A}\bv_{\eta \to -0} =
- 16 \pi\,\chi \; ,
\lab{israel-junction}
\ee
which yields the following relation between the Reissner-Nordstr{\"o}m
parameters and the dynamical \textsl{LL}-brane tension:
\be
4\pi\chi\,r_0^2 + r_0 - M = 0 \quad, \;\;\mathrm{where} \;\;
r_0 = M +\sqrt{M^2 -e^2} \; .
\lab{parameter-matching}
\ee
Eq.\rf{parameter-matching} indicates that the dynamical brane tension must
be {\em negative}. Eq.\rf{parameter-matching} reduces to a cubic equation
for the Reissner-Nordstr{\"o}m mass $M$ as function of $|\chi|$:
\be
\bigl( 16\pi\,|\chi|\, M - 1\bigr) \( M^2 - e^2\) + 16\pi^2 \chi^2 e^4 = 0 \; .
\lab{M-RN}
\ee
In the special case of Schwarzschild wormhole ($e^2 = 0$) the Schwazrschild mass
becomes:
\be
M = \frac{1}{16\pi\,|\chi|} \;\; .
\lab{M-Schw}
\ee
Notice that, for large values of the 
\textsl{LL-brane} tension $|\chi|$, $M$ is very small. In particular, 
$M << M_{Pl}$ for $|\chi| > M_{Pl}^3$ ($M_{Pl}$ being the Planck mass).

\section{Conclusions}
\label{conclude}
In the present note we have constructed a wormhole solution by sewing together
two copies of exterior Reissner-Nordstr{\"o}m space-time regions at a 
Reissner-Nordstr{\"o}m outer
horizon via a \textsl{LL-brane} with a negative dynamical tension. This
\textsl{LL-brane} provides a theoretically sound {\em non-phenomenological}
gravitational source for the Reissner-Nordstr{\"o}m wormhole since its dynamics, in
particular its surface energy-momentum tensor, are derived from a
well-defined world-volume Lagrangian action \rf{LL-brane}. Furthermore, 
let us stress that the \textsl{LL-brane} is electrically neutral and at the same 
time the Reissner-Nordstr{\"o}m wormhole appears to possess two oppositely charged 
sources -- one for each Reissner-Nordstr{\"o}m region beyond the common horizon.

According to Eq.\rf{parameter-matching} (in particular Eq.\rf{M-Schw}) wormholes 
built from \textsl{LL-branes} with very high negative tension have a small mass. 
To this end it is interesting to note that one can obtain baby universe solutions at 
very small energy cost by considering high surface tensions too, which similarly 
require a wormhole, although there the tension is positive (refs.\ct{baby-univ}).
Notice however that in refs.\ct{baby-univ} the solutions are time-dependent and
suffer from a singular initial problem. In the present work there is no singularity
-- the would-be singularities have been ``surgically removed'' by the wormhole
matching.

On the other hand, for small values of the \textsl{LL-brane} tension $|\chi|$
Eq.\rf{parameter-matching} implies that the Reissner-Nordstr{\"o}m geometry of the
wormhole must be near extremal ($M^2 \simeq e^2$).


\textbf{Acknowledgments.}
E.N. and S.P. are supported by European RTN network
{\em ``Forces-Universe''} (contract No.\textsl{MRTN-CT-2004-005104}).
They also received partial support from Bulgarian NSF grants \textsl{F-1412/04}
and \textsl{ID01/133}.
Finally, all of us acknowledge support of our collaboration through the exchange
agreement between the Ben-Gurion University of the Negev (Beer-Sheva, Israel) and
the Bulgarian Academy of Sciences.


\end{document}